\documentclass{aa}  
\usepackage{graphicx}
\usepackage{txfonts}
\usepackage{hyperref}
\usepackage{subfigure}
\usepackage{afterpage}
\usepackage{comment}
\usepackage{layouts}
\usepackage{multirow}
\usepackage{xcolor}
\usepackage{listings} 
\usepackage{amsmath,amssymb}
\usepackage{float}

\definecolor{codegreen}{rgb}{0,0.6,0}
\definecolor{codegray}{rgb}{0.5,0.5,0.5}
\definecolor{codepurple}{rgb}{0.58,0,0.82}
\definecolor{backcolour}{rgb}{0.95,0.95,0.92}

\lstdefinestyle{mystyle}{
    backgroundcolor=\color{backcolour},   
    commentstyle=\color{codegreen},
    keywordstyle=\color{magenta},
    numberstyle=\tiny\color{codegray},
    stringstyle=\color{codepurple},
    basicstyle=\ttfamily\footnotesize,
    breakatwhitespace=false,         
    breaklines=true,                 
    captionpos=b,                    
    keepspaces=true,                 
    numbers=none,                    
    numbersep=5pt,                  
    showspaces=false,                
    showstringspaces=false,
    showtabs=false,                  
    tabsize=2
}

\newcommand{\zerodel}{.\kern-\nulldelimiterspace}

\newcommand{\SD}[1]{\textcolor{purple}{#1}}

\begin{document} 

   \title{\texttt{jaxspec} : a fast and robust \texttt{Python} library for X-ray spectral fitting} 

   \author{
        S. Dupourqué \inst{\ref{inst1}}
        \and 
        D. Barret \inst{\ref{inst1}}
        \and
        C. M. Diez \inst{\ref{inst2}}
        \and 
        S. Guillot \inst{\ref{inst1}}
        \and 
        E. Quintin \inst{\ref{inst1},\ref{inst2}}
    }

   \institute{IRAP, Université de Toulouse, CNRS, CNES, UT3-PS, Toulouse, France \\ \email{sdupourque@irap.omp.eu}\label{inst1} 
   \and European Space Agency (ESA), European Space Astronomy Centre (ESAC), Camino Bajo del Castillo s/n, 28692 Villanueva de la Cañada, Madrid, Spain\label{inst2} 
}

   \date{Received ???; accepted ???}

 
  \abstract{Inferring spectral parameters from X-ray data is one of the cornerstones of high-energy astrophysics, and is achieved using software stacks that have been developed over the last twenty years and more. However, as models get more complex and spectra reach higher resolutions, these established software solutions become more feature-heavy, difficult to maintain and less efficient.
  }{
  We present \texttt{jaxspec}, a \texttt{Python} package for performing this task quickly and robustly in a fully Bayesian framework. Based on the \texttt{JAX} ecosystem, \texttt{jaxspec} allows the generation of differentiable likelihood functions compilable on core or graphical process units (resp. CPU and GPU), enabling the use of robust algorithms for Bayesian inference.
  }{
  We demonstrate the effectiveness of \texttt{jaxspec} samplers, in particular the No U-Turn Sampler, using a composite model and comparing what we obtain with the existing frameworks. We also demonstrate its ability to process high-resolution spectroscopy data and using original methods, by reproducing the results of the \textit{Hitomi} collaboration on the Perseus cluster, while solving the inference problem using variational inference on a GPU.
  }{
  We obtain identical results when compared to other softwares and approaches, meaning that \texttt{jaxspec} provides reliable results while being $\sim 10$ times faster than existing alternatives. In addition, we show that variational inference can produce convincing results even on high-resolution data in less than 10 minutes on a GPU.  
  }{
  With this package, we aim to pursue the goal of opening up X-ray spectroscopy to the existing ecosystem of machine learning and Bayesian inference, enabling researchers to apply new methods to solve increasingly complex problems in the best possible way. Our long-term ambition is the scientific exploitation of the data from the \textit{newAthena} X-ray Integral Field Unit (X-IFU).
  }


   \maketitle
%

\section{Introduction}

Spectral fitting is one of the cornerstones of X-ray astrophysics and has been widely used and developed since the 1990s, as the number of X-ray instruments has increased in the last decades. An X-ray spectrum, represented by a number of counts in energy bins, can be reduced to meaningful parameters using an appropriate physical model. 
Constraining the plausible values of these parameters given an X-ray spectrum is usually achieved using publicly available software, such as \texttt{XSPEC} \citep{arnaud_xspec_1996}, \texttt{SPEX} \citep{kaastra_spex_1996}, \texttt{sherpa} \citep{freeman_sherpa_2001, doe_developing_2007}, or \texttt{ISIS} \citep{houck_isis_2000}. 

The typical approach is to define a fit statistic (e.g. $\chi^2$ or C-stat, see \citealt{cash_parameter_1979}) and optimise it using gradient-free minimisation algorithms. This type of approach is efficient if the initial parameter estimates are close to their optimal values, but is highly dependent on the dimensionality and smoothness of the parameter space. In general, there is no guarantee that these algorithms will not get stuck in local minima, which would ultimately lead to incorrect physical interpretations. Moreover, deriving a meaningful error budget in this situation is not trivial and often requires an exhaustive evaluation of the fit statistic in an N-dimensional lattice in the neighbourhood of the best fit, which is not efficient as the number of parameters N increases. 

On the other hand, Bayesian approaches provide a sample of parameters that are distributed according to the likelihood. The sample of posterior parameters carries the information about the dispersion and is less likely to get stuck in a local minimum \citep[e.g. ][]{bonson_how_2016, choudhury_testing_2017, barret_inferring_2019}. This type of approach requires orders of magnitude more computation time than direct minimisation. Posterior samples can also be generated using Markov Chain Monte Carlo (MCMC) approaches, or other algorithms such as nested sampling implemented in \texttt{BXA} \citep{buchner_x-ray_2014, buchner_ultranest_2021}, which can efficiently check their convergence. More recently, we proposed a new approach using neural networks to directly learn the posterior distribution, given a set of simulated spectra \citep[see \texttt{SIXSA}, ][]{barret_simulation-based_2024}. This simulation-based inference approach showed similar performance compared to the existing solutions, while being much faster.

As a more general trend, we have seen major advances in the field of machine learning over the last decade, mainly based on the use of back-propagation algorithms \citep{rumelhart_learning_1986}. Today's neural networks can be parameterised with up to billions of parameters, and finding the best parameters to achieve a given task is unthinkable using classical minimisation algorithms. The back-propagation algorithm allows the gradient of any function to be estimated, as long as it can be represented by a computational graph of differentiable components. This additional gradient information allows the use of powerful minimisation algorithms, such as the \texttt{adam} \citep{kingma_adam_2017} algorithm, which can find optimal values for parameters in high-dimensional spaces. Modern frameworks such as \texttt{tensorflow} \citep{abadi_tensorflow_2015, abadi_tensorflow_2015-1}, \texttt{PyTorch} \citep{paszke_pytorch_2019} or \texttt{JAX} \citep{bradbury_jax_2018} can provide this gradient information automatically. In particular, \texttt{JAX} is already used in the astrophysics community to facilitate the inference of cosmological parameters \citep[e.g.][]{campagne_jax-cosmo_2023, piras_cosmopower-jax_2023, balkenhol_candl_2024} by providing differentiable likelihood functions.

In this article, we present \texttt{jaxspec}, a \texttt{JAX}-based X-ray spectral fitting package written in pure \texttt{Python}. Using differentiable programming, \texttt{jaxspec} uses gradient-based MCMC samplers and variational inference algorithms to quickly and efficiently provide distributions for the spectral model parameters. We provide a description of how \texttt{jaxspec} is designed, and we benchmark it against existing packages to demonstrate the benefits of just-in-time compilation and auto-differentiability. By studying \textit{Hitomi}/Soft X-ray Spectrometer (SXS) observations of the core of the Perseus cluster and constraining the gas motions using a Bayesian hierarchical model, we demonstrate the feasibility of variational inference for high-resolution X-ray spectral fitting. Finally, we discuss the philosophy behind the development of \texttt{jaxspec} and our desire to provide software that is modern, flexible, open source, and driven by the community.

\section{Description of \texttt{jaxspec}}
\label{sec:description}

\begin{figure*}
    \centering
    \subfigure{
\includegraphics[height=0.35\hsize]{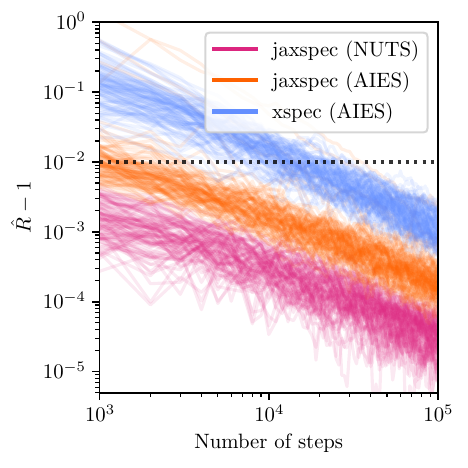}}
\subfigure{
\includegraphics[height=0.35\hsize]{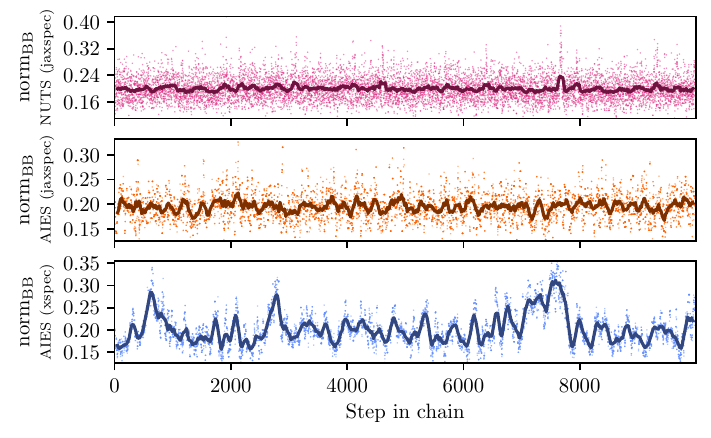}}
    \caption{(\textit{left}) $\hat{R}$ statistic computed for 100 runs of $10^5$ steps of NUTS and AIES using the \texttt{jaxpsec} and \texttt{xspec} implementations. The usual convergence threshold of $\hat{R} < 1.01$ is shown in black. (\textit{right}) Value of the parameter $\text{norm}_{\text{BB}}$ for $10^4$ steps of a single chain in the sample using NUTS and AIES with \texttt{jaxpsec} and \texttt{xspec} implementation. The solid line represents a 100 step moving average of the parameter value.}
    \label{fig:mcmc_benchmark}
\end{figure*}

We have designed \texttt{jaxspec} as a software package written entirely in \texttt{Python}. By leveraging key packages from the data science ecosystem, the source code itself is greatly simplified compared to other existing solutions, while ensuring some sustainability over time. In particular, \texttt{jaxspec} relies on two widely used and maintained dependencies: . \texttt{JAX} \citep{bradbury_jax_2018}, a library similar to \texttt{tensorflow} and \texttt{pytorch} for writing automatically differentiable code that can be compiled at runtime. \texttt{JAX} is widely used in the machine learning community and allows code to be executed transparently on core, graphical or tensor process units (resp. CPU, GPU and TPU) to achieve significant speed-up; ii. \texttt{numpyro} \citep{phan_composable_2019}, a probabilistic programming library that implements the construction of hierarchical Bayesian models and the tools needed to perform the associated inference. In particular, \texttt{numpyro} includes a \texttt{JAX} implementation of the No U-Turn sampler \citep{hoffman_no-u-turn_2014} and an interface for applying variational inference methods. We are distributing \texttt{jaxspec} to the community as a \texttt{Python} package that can be downloaded and installed directly from the \texttt{pypi} portal. The codebase is open source and publicly available in a GitHub repository\footnote{\url{https://github.com/renecotyfanboy/jaxspec}} and documented\footnote{\url{http://jaxspec.rtfd.io/}}.

Spectral models in \texttt{jaxspec} can be built in the same way as one would expect from \texttt{xspec}, i.e.  by organically nesting model components. We have already implemented some commonly used and analytically computable model components\footnote{Implemented models are available at~\url{https://jaxspec.rtfd.io/en/stable/references/model/}} (e.g. blackbody spectrum, power-law spectrum) and tabulated absorption models, such as the Tübingen-Boulder Interstellar Medium absorption model \texttt{tbabs} \citep{wilms_absorption_2000}. \texttt{jaxspec} does not (yet) implement many of the models available in \texttt{xspec}, and porting legacy \texttt{Fortran} / \texttt{C} models to pure \texttt{JAX} is neither a quick nor an easy task. However, models that are interpolated on rectangular grids are readily implementable in \texttt{jaxspec}, such as the \texttt{nsatmos} neutron star atmospher model \citep{heinke_hydrogen_2006}. As described later in Sect.~\ref{sec:discussion}, we see this task as a joint effort by the community to update, maintain and share its models, and will do our best to help model developers integrate their work into this codebase.

These models are folded into the instrumental response using matrix multiplications, which allow the expected number of counts $\vec{\lambda}_i$ in each instrumental bin to be estimated, given a set of parameters $\vec{\theta}$ and for a particular observing setup. Each spectral model is parameterised by a subset of the parameters $\vec{\theta}$. Finding the best set of parameters to describe the observed data requires a metric to define how good a model is. In the case of X-ray spectral fitting, the detection of photons within instrumental bins can be described as a counting process, which means it can be modelled as a Poisson random variable. The corresponding log-likelihood\footnote{Note that in Eq.~\ref{eq:likelihood} we write $\Gamma(c_i +1)$ instead of $c_i!$. The numerical evaluation of Euler $\Gamma$ can be much faster and easier to differentiate. However, since this term does not vary for a given inference problem, its value is cached during the sampling.} for the expected photons $\vec{\lambda} (\vec{\theta})$ according to the model and the observed values $\vec{c}$ is written as: 

\begin{equation}
    \log \mathcal{L}(\vec{c}|\vec{\theta}) = \sum_i c_i \log \lambda_i(\vec{\theta}) - \log \Gamma(c_i +1) - \lambda_i(\vec{\theta})
    \label{eq:likelihood}
\end{equation}

\noindent where $\sum_i$ denotes the summation over the instrument bins. This Poisson likelihood is directly related to the well-known Cash statistic \citep{cash_parameter_1979}, which is actually a transformation of the Poisson likelihood that corresponds to the $\chi^2$ statistic at high counts. With \texttt{jaxspec} we rely solely on the Poisson likelihood, as the $\chi^2$ likelihood has historically been used to study high count spectra and to simplify error calculations, and is biased at lower counts \citep[see, e.g. ][]{humphrey_2_2009, kaastra_use_2017, buchner_statistical_2023}. Finding the best set of parameters that maximises this likelihood and its associated variance can be achieved using Bayesian inference approaches.

\section{Benchmarking Bayesian inference in \texttt{jaxspec}}

Solving an inference problem with Bayesian approaches requires inverting this likelihood into a posterior distribution of $\vec{\theta}$ constrained to the observed number of counts $\vec{c}$. This requires defining prior distributions for $\vec{\theta}$ that encapsulate our (lack of) knowledge about the expected values of our model parameters. Once the Bayesian inference problem is set up, the easiest way to solve it is to sample the parameter values $\left\{\vec{\theta}\right\}_i$ from the posterior distribution. There are many methods to do this, such as MCMC, nested sampling, or variational inference. 

\subsection{Quality of MCMC samples}

MCMC sampling has been one of the most widely used approaches to obtaining parameter samples that are distributed according to the posterior distribution of a Bayesian inference problem since the Metropolis Hastings algorithm was proposed \citep{hastings_monte_1970}. MCMC algorithms generate chains of samples by moving gradually through the parameter space, making small, local changes at each step. They are designed so that the stationary distribution of these chains is the posterior distribution of the parameters. In this sense, reaching the stationary distribution requires a sufficiently large number of iterations, which is not trivial to quantify. Many solutions based on MCMC sampling can provide posterior samples. The Affine Invariant Ensemble Sampler \citep[AIES, ][]{goodman_ensemble_2010} uses multiple chains that are not independent (often referred to as walkers) to shift the proposal distribution towards optimal regions of the parameter space. This is the sampler implemented in the well-known \texttt{emcee} package \citep{foreman-mackey_emcee_2013}, and also the default sampler implemented in \texttt{xspec}. While the traditional MCMC sampling algorithm requires numerous steps to achieve stationarity, gradient-based samplers can significantly reduce the number of steps required. Hamiltonian Monte Carlo \citep[e.g. ][]{betancourt_hamiltonian_2013} uses the gradient of the likelihood to perform “energy saving” steps in the parameter space, allowing much better exploration and faster convergence. In particular, the No U-Turn sampler \citep[NUTS, ][]{hoffman_no-u-turn_2014} uses this dynamic to adaptively explore the likelihood, setting the step size depending on the local curvature. Since \texttt{jaxspec} provides auto-differentiable likelihoods, we use the \texttt{numpyro} implementation of NUTS as the default sampler, and also compare the \texttt{numpyro} and \texttt{xspec} implementations of AIES.

To compare the performance of \texttt{jaxspec} with existing solutions, we chose to solve a similar inference problem using different methods. The task is to obtain the posterior distribution of the five parameters of an absorbed dual-component model (\texttt{tbabs*(powerlaw+blackbodyrad))} fitted to an observation of NGC 7793 ULX-4. Ultra-luminous X-ray sources \citep[ULXs, e.g.,][]{kaaret_ultraluminous_2017} are highly-accreting stellar-mass binary systems, in which the precise physical processes at play are still largely debated. In particular, some ULXs display a two-component spectral emission in X-rays \citep[e.g., ][]{sutton_ultraluminous_2013}, one component (thermal) being generally interpreted as an accretion disk and the second (power-law-like) as either a hot accretion column or a Compton-scattering outflow \citep[e.g., ][]{gurpide_long-term_2021}. The relative temporal evolution of one component with respect to the other is generally different in ULXs compared to standard X-ray binaries \citep[e.g., ][]{koliopanos_ulx_2017}. As such, precisely and carefully constraining the parameters from both components can shed some light on the physics of these objects. We use the \textit{XMM-Newton} observations of NGC 7793 ULX-4 as reduced in \cite{quintin_new_2021}, and focus on the spectrum produced by the EPIC-PN instrument, with no background. All calculations in this section are performed on a single core in order to obtain comparable results between the different methods used, but it is worth noting that they all benefit from being performed on multiple cores.

In this benchmark, we run MCMC chains using AIES and NUTS and measure how quickly the chain converges to a stationary distribution using the $\hat{R}$ statistic \citep{vehtari_rank-normalization_2021} computed for an increasing number of samples. The $\hat{R}$ statistic compares the intra- and inter-chain variance to assess their individual convergence.  AIES is run with $2^5 = 32$ walkers, as suggested by the number of parameters in our model, while NUTS is run with 4 chains. We run $10^3$ burn-in steps and $10^5$ sampling steps, and compute the $\hat{R}$ statistic for an increasing number of samples, with a step of 1000. The $\hat{R}$ statistic is computed using only the best 4 chains of each sampler, defined in terms of the best likelihood on average. We show in the left panel of Fig.~\ref{fig:mcmc_benchmark} that the $\hat{R}$ decreases as the number of steps increases, as the chains begin to converge. The usual convergence threshold of $\hat{R} < 1.01$ is reached well within the first $1000$ steps with NUTS, $\sim 5000$ steps with AIES in \texttt{jaxspec} and $25000$ steps with AIES in \texttt{xspec}. The right panel of Fig.~\ref{fig:mcmc_benchmark} shows the individual samples of a single chain using the $\text{norm}_{\text{BB}}$ parameter. As expected, we observe that NUTS is much better at providing uncorrelated samples, while AIES and especially the \texttt{xspec} implementation provide samples with visible correlations, meaning that it takes a large number of steps and additional thinning to get a meaningful sample of parameters from the posterior distribution.

\subsection{Comparison of inference time for various methods}

Another common approach to solving Bayesian inference problems is to use nested sampling \citep[NS, see, e.g.  ][]{buchner_nested_2023}, which is a powerful alternative to MCMC sampling. NS relies on a population of points in the parameter space that evolves according to their likelihood at each step. As the number of steps increases, the points with the lowest likelihood are excluded and new points are drawn according to the distribution of the remaining points. Such evolutionary approaches shrink the volume occupied by the samples until there is no further improvement in likelihood. Unlike MCMC approaches, NS provides a stopping criterion and can explore a much larger part of the parameter space, providing robust estimates for the posterior distribution, especially in multimodal cases. A popular implementation of NS used in X-ray spectral fitting can be found in \texttt{BXA} \citep{buchner_x-ray_2014}, which interfaces the NS algorithm of \texttt{ultranest} \citep{buchner_ultranest_2021} with \texttt{xspec} or \texttt{sherpa}. Throughout this paper we will only use the \texttt{xspec} interface.

\begin{figure}[t]
    \centering
    \includegraphics[height=\hsize]{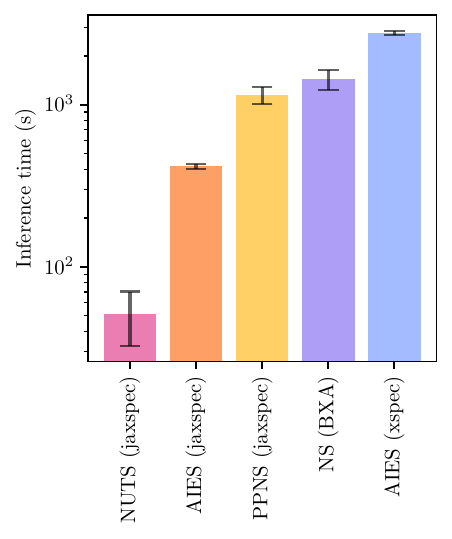}
    \caption{Time per run and standard deviation for NUTS, AIES and NS in both \texttt{jaxspec}, \texttt{BXA} and \texttt{xspec}, estimated for 100 runs of each algorithm on a 5-parameter, 3-component inference problem.}
    \label{fig:time_benchmark}
\end{figure}

In this subsection, we use different inference algorithms and compare their results and total inference time. We use the previously introduced NUTS and AIES samplers implemented in \texttt{jaxspec} and \texttt{xpsec}. We also compare the existing \texttt{BXA} sampler with the implementation of phantom-powered nested sampling \citep[PPNS, ][]{albert_phantom-powered_2023} provided in \texttt{jaxns}. We use the same observational setup as in the previous subsection and fit the same 5 parameter model. We run all algorithms with default hyperparameters for sampling and termination, and use the previously determined number of steps to convergence for the MCMC sampler. 

\begin{figure*}[t]
    \centering
    \includegraphics[height=0.65\hsize]{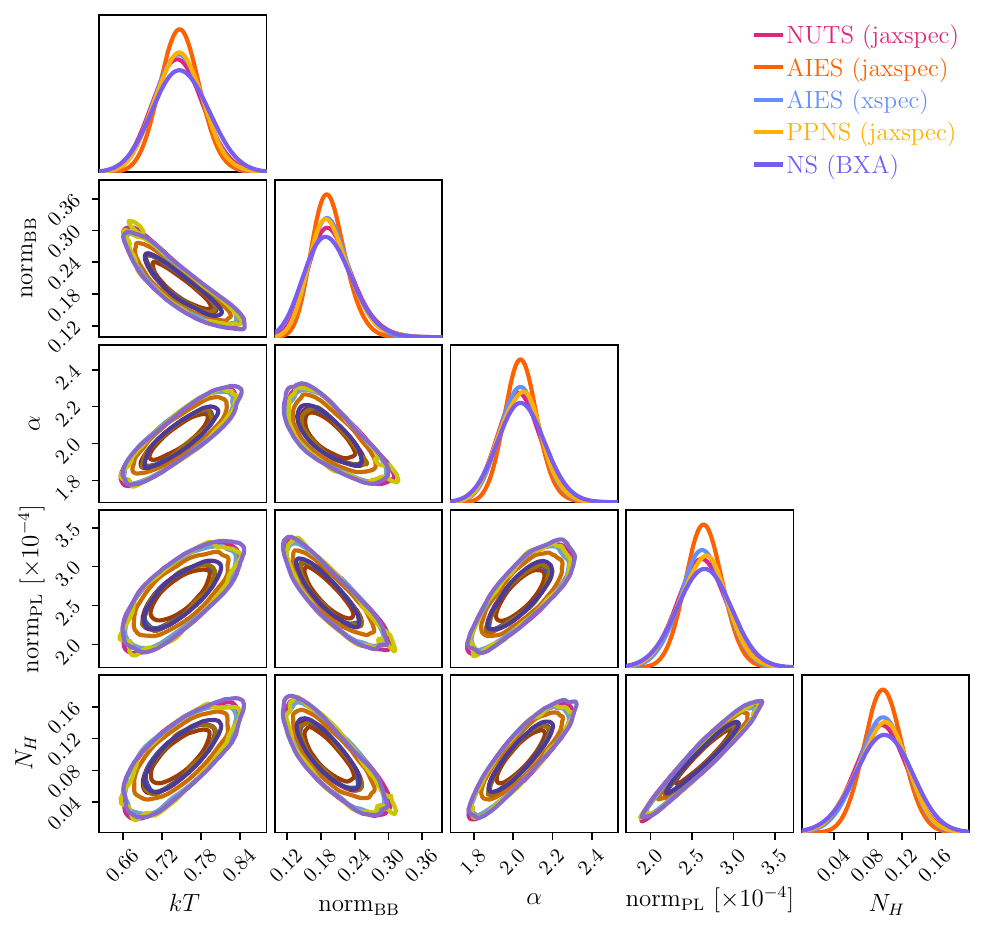}
    \caption{Posterior distribution of \texttt{tbabs*(powerlaw+blackbodyrad)} parameters from this benchmark, drawn with the $10^4$ last steps of a run of NUTS and AIES using the \texttt{jaxpsec} and \texttt{xspec} implementations, and as provided by the NS algorithms implemented in \texttt{BXA} and \texttt{jaxspec}.}
    \label{fig:mcmc_corner}
\end{figure*}

The time for 100 runs of each inference approach is recorded and plotted in Fig.~\ref{fig:time_benchmark}. For this particular case, we observe that using NUTS is an order of magnitude faster than the other approaches, AIES in \texttt{jaxspec} is slightly faster, and PPNS in \texttt{jaxspec} is equivalent to NS in \texttt{BXA} in terms of speed. For a well-behaved inference problem, NUTS can generate stable posterior distributions in less than a minute. In Fig.~\ref{fig:mcmc_corner}, we show the results of a single run of each algorithm. The posterior distributions are in good agreement. We note that the AIES implementation in \texttt{numpyro}, which is used in \texttt{jaxspec}, produces marginal distributions that are tighter than the other algorithms, due to differences in the underlying implementations. In a nutshell, the \texttt{xspec} implementation uses “stretch” moves from \cite{goodman_ensemble_2010} while \texttt{jaxspec} uses “differential evolution” moves from \cite{nelson_run_2014}. The second converges faster but struggles to explore the tail of the distribution, slightly underestimating the spread of the parameters. Increasing the number of walkers should solve this. In either case, the user can choose to use the “stretch” move or a combination of the two in \texttt{numpyro} to achieve the desired behaviour. A composite model like the one presented in this benchmark is a standard use case in the community, making this benchmark representative of a typical inference case for accreting objects. As it is easy to choose a specific sampler in \texttt{jaxspec}, the user can try different approaches to solve a single problem, leading to robust results in the end.

\section{High-resolution spectroscopy with \texttt{jaxspec}}
\label{sec:hitomi}
For the sake of demonstrating the capabilities of \texttt{jaxspec}, we consider the high-resolution X-ray spectrum from the \textit{Hitomi}/SXS observations of the Perseus core, obtained with the exquisite spectral resolution of $\sim 5 $eV, which enabled the first high-precision direct measurement of the motions of the intracluster gas \citep{hitomi_collaboration_quiescent_2016, hitomi_collaboration_atmospheric_2018}. By studying the Fe xxv He$\alpha$, He$\beta$ and He$\gamma$ complexes, the bulk and velocity dispersions have been constrained to $\sim 50$ km/s and $\sim 200$ km/s respectively in the central regions of the cluster. The high spectral resolution results in extremely dense spectra with large response matrices, which makes direct minimisation more difficult and the general inference more computationally expensive within existing frameworks. We propose to demonstrate the use of \texttt{jaxspec} to accurately model the observations of the Perseus core with \textit{Hitomi}. To achieve this, we have used the publicly available data, bearing in mind that these data are subject to calibration problems that limit the scientific scope of the results. We use the pipeline products from \texttt{ObsIds} 40, 50 and 60. The spectra are optimally binned using the algorithm from \cite{kaastra_optimal_2016}. As in \cite{hitomi_collaboration_quiescent_2016}, we focus on modelling the Fe XXV He$\alpha$ complex with Gaussian emission lines of known energies in the [6.49, 6.61] keV band. The measurement of the centroid shift and broadening of these lines provides information about the dynamics of the intra-cluster medium along the line of sight, and is a direct measure of the turbulent motions. In Appendix ~\ref{app:perseus-model}, we define a spectral model consisting of nine Gaussian emission lines with known energies to assess the centroid shift and broadening resulting from the gas motion. We use the cluster redshift determination from \cite{hitomi_collaboration_atmospheric_2018} and assume a Gaussian prior such as $z \sim \mathcal{N}(0.017284, 0.00005)$. All the model parameters are constrained using the three observations simultaneously.

\begin{figure*}[!t]
\centering
\subfigure[ELBO during minimisation]{
\includegraphics[width=0.45\hsize]{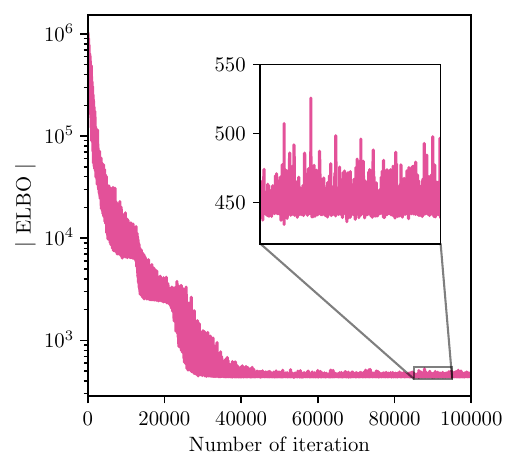}}
\subfigure[Posterior distributions]{
\includegraphics[width=0.45\hsize]{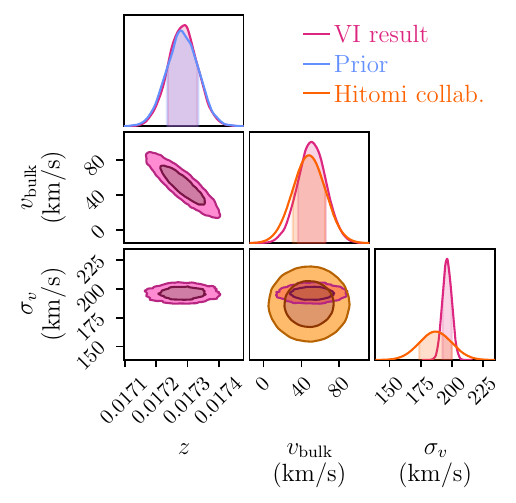}}
\subfigure[Posterior predictive on observed spectra]{
\includegraphics[width=0.85\hsize]{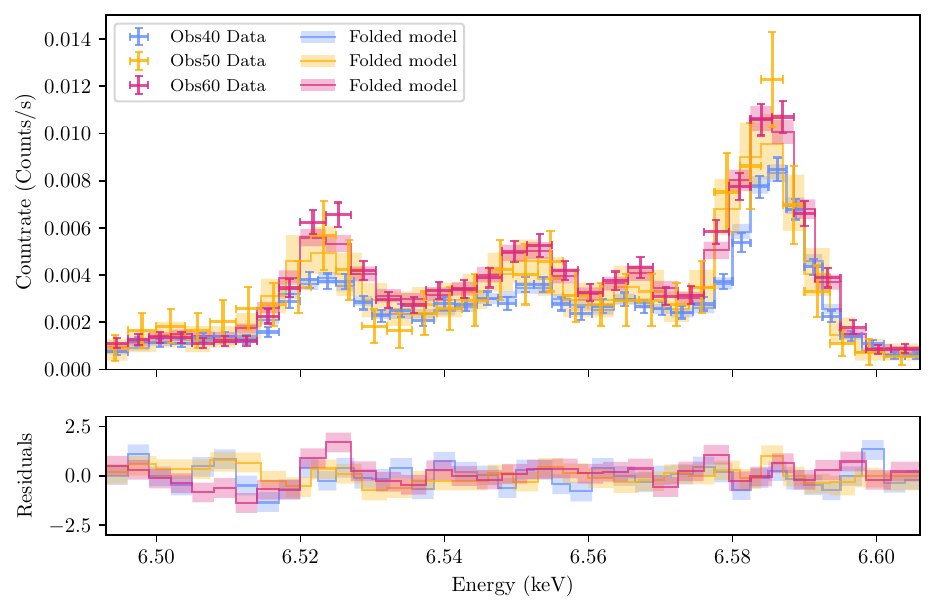}}
        \caption{(a) ELBO metric during variational distribution optimisation. (b) Redshift, bulk motion $v_{\text{bulk}}$ and velocity dispersion $\sigma_{v}$ obtained for the central regions of the Perseus cluster as seen by \textit{Hitomi}/SXS, superimposed with the prior distribution for the redshift and the previous results from \citep{hitomi_collaboration_quiescent_2016} in the central region. (c) The count spectra of the observations used in this spectral fit, superimposed with the folded model for each observation, and the corresponding residuals in $\sigma$.}
    \label{fig:hitomi-broadening}
\end{figure*}

As a novelty, we solve this Bayesian inference problem using a variational inference (VI) approach. This family of methods takes a different approach from the MCMC family. Instead of sampling directly from the posterior distribution, it analytically computes an approximation to the posterior density. It requires the specification of a family of distributions $q(\theta)$, called the variational distribution, which is fitted to the true posterior distribution. In practice, this is equivalent to saying that the posterior distribution would be well approximated for example by a Gaussian distribution for each variable, a multivariate Gaussian for all parameters with a mean and a correlation matrix to be determined, or any analytically computable and parameterised probability density. Selecting a surrogate posterior distribution from a set of known densities transforms the inference problem into an optimisation problem in which we try to recover the best parameters of the surrogate distribution. This in turn requires defining a distance between the variational distribution and the true posterior, minimized by adjusting the variational distribution parameters. One commonly used measure between distributions in machine learning studies is the Kullback-Leibler divergence \citep[KL divergence, ][]{kullback_information_1951}. It shows how much information is lost when a model distribution is used to approximate the true underlying distribution. For the optimisation process in VI, the Evidence Lower Bound is the commonly used metric, which is directly derived from the KL divergence between the variational distribution and the posterior distribution. For further insight, we direct the interested reader to the review by \cite{blei_variational_2017}. The ELBO is defined as the difference between the expectation (with respect to the variational distribution $q(\vec{\theta})$) of the logarithm of the joint distribution $p\left( \vec{\theta}, \vec{X} \right)$ and the logarithm of the variational distribution:

\begin{equation}
    \text{ELBO}(q(\vec{\theta})) = \mathbb{E}_{q(\vec{\theta})} \left\{ \log p\left( \vec{\theta}, \vec{X} \right) \right \} - \mathbb{E}_{q(\vec{\theta})} \left\{ \log q\left(\vec{\theta}\right) \right \},
\end{equation}

\noindent while the best surrogate posterior distribution $q^*(\vec{\theta})$ is the one that maximises the ELBO:

\begin{equation}
    q^*(\vec{\theta}) = \underset{q(\vec{\theta})}{\text{argmin}} \: \left\{-\text{ELBO}(q(\vec{\theta}))\right\}.
\end{equation}

We aim to demonstrate the feasibility and effectiveness of VI for X-ray spectral fitting. To our knowledge, this is the first time that such methods have been used for X-ray spectroscopy in astrophysics, with only a handful of works on VI in astrophysics in general \citep[e.g. ][]{gunapati_variational_2022}. It has to be noted that classical approaches can also work here. We assume a multivariate Gaussian distribution as the variational distribution for these observations. This optimisation problem takes particular advantage of the differentiability of its metric, provided here by the auto-differentiation in \texttt{JAX}. This makes it possible to mobilise first-order optimisers \citep[e.g. \texttt{adam}, ][]{kingma_adam_2017} using gradient information to find a satisfactory minimum for the problem. These algorithms require the user to set hyperparameters, such as the learning rate, which defines the length of each step in the parameter space, or the total number of iterations. For this particular problem, we use the \texttt{adamw} optimiser \citep{loshchilov_decoupled_2019} with default parameters and a learning rate of $5^{-4}$ for $10^5$ iterations. The gradient descent can be run on the GPU and takes $\sim 7$ minutes on an Nvidia A2. In Fig.~\ref{fig:hitomi-broadening} we show the ELBO evolution for the $10^5$ \texttt{adamw} steps, the resulting redshift, bulk motion and velocity dispersion in the observer frame, and the distribution of the posterior folded spectra. The stationarity of ELBO, coupled with the small residuals on the observed spectra, indicates a qualitative fit. The line-of-sight bulk motion of $51 \pm 15$ km/s and the velocity dispersion of $196 \pm 4$ km/s are compatible with those measured by \cite{hitomi_collaboration_quiescent_2016} in the central region of Perseus, but we stress that our results are not directly comparable, as our analysis is averaged over all Perseus observations and all SXS pixels, and a more exhaustive treatment including systematic uncertainties and calibration issues is beyond the scope of this paper. We also show that the posterior has the expected behaviour, as no insight into the cluster redshift is gained due to a degeneracy with the bulk motion captured by our variational distribution. This confirms the efficiency of VI methods for X-ray spectral fitting.

\section{Enabling a community-driven package}
\label{sec:discussion}
As the X-ray spectral fitting ecosystem is already relatively crowded, we feel it is important to justify our efforts to add a new library to it. Beyond the efficiency of \texttt{JAX} and the complexity of scientific use cases enabled by the implementation of state-of-the-art sampling algorithms, the key strengths of \texttt{jaxspec} are its ease of installation and use. It is worth noting that our long-term goal is to exploit the scientific data produced by high-resolution spectrometers such as \textit{XRISM}/Resolve and the future \textit{newAthena}/XIFU \citep{barret_athena_2023}, which would greatly benefit from modern and fast spectral fitting softwares.

\subsection{From the casual user point of view}

The objective is to provide users with a straightforward software solution for spectral inference that is both readily deployable and intuitive to use. As \texttt{jaxspec} is not dependent on \texttt{xspec} or any equivalent, the installation process is as straightforward as that of any \texttt{Python} package, requiring only the \texttt{pip} command. The software is fully documented and provides examples of inference. As detailed in the Appendix~\ref{app:usecase}, the syntax for building models and fitting them to observations is kept as simple as possible. For scientists using other tools, the interface we propose should be familiar. The default set-ups for inference are fast and efficient at providing reliable scientific results. Nevertheless, as \texttt{jaxspec} exposes its likelihoods, the user can extend its use beyond what is provided in the package.

\subsection{From the contributor point of view}

The \texttt{jaxspec} codebase represents a fresh start from the existing frameworks. It greatly reduces the code complexity when compared to existing alternatives. Relying on the \texttt{Python} ecosystem for data science reduces the need for maintaining heavy boilerplates. The continuous integration framework, with automated tests, code linters and automated documentation, allows deploying updates and patches easily. The package is open source and hosted on \texttt{GitHub}, and individual contributions are encouraged. This facilitates a collaborative environment in which users can raise issues and directly contribute to the codebase by correcting and extending source code. In particular, it is possible to extend the spectral models, as was previously possible with \texttt{xspec} and other similar software. Analytical models and models interpolated on regular grids can be readily implemented.

\section{Conclusions}

In this paper, we presented \texttt{jaxspec}, a pure \texttt{Python} and \texttt{JAX} software for X-ray spectral fitting. Since \texttt{JAX} is differentiable and compilable, powerful samplers such as the No U-Turn sampler can be used to solve the inference problems quickly and reliably, running seamlessly on CPU or GPU. We show that it outperforms existing software by an order of magnitude in execution time on a 5-parameter, 3-component model, while providing robust posterior distributions. The compilable and auto-differentiable likelihood provided by \texttt{jaxspec} allows the use of variational inference, which can provide accurate results for high dimensional models, as demonstrated on high-energy resolution \textit{Hitomi}/SXS spectra. The likelihood functions built into \texttt{jaxspec} are exposed so that it can be interfaced with external packages such as \texttt{ultranest}, giving users many options.

The current limitation of \texttt{jaxspec} is the library of spectral models that can be fitted, as models must be translated in pure \texttt{JAX}. Regularly tabulated models are already easy to port. We also plan to integrate existing features from other software, such as direct minimisation, multi-instrument gain and shift, convolutional spectral components, restriction prior as implemented in \texttt{SIXSA} etc. and to systematically implement \texttt{xspec} models, which could be used with non-differentiable methods. We also plan to explore the possibility of using neural networks to create surrogate models to port existing models to our code base. If successful, this approach would have the advantage of providing naturally differentiable codes and fast to run on CPUs and GPUs, at the cost of controlled error in model evaluation (see e.g. \cite{varma_surrogate_2019, varma_surrogate_2019-1} for gravitational waves or \cite{euclidcollaboration_euclid_2021} for cosmology applications).  As we have taken great care in building this software using good software engineering practices, with a contribution policy, automated testing, and code quality linters, we hope that it will evolve according to the needs of the community, encouraging individual contributions to the codebase and providing astrophysicists with a simple yet reliable tool for X-ray spectral fitting.  


\begin{acknowledgements}
SD and DB would like to thank Florent Castellani, and Alexeï Molin for their early contributions to this project, and Fabio Acero for the fruitful discussions about spectral fitting. SD and DB thank the Centre National d’Etudes Spatiales for financial support. All the calculations involving GPUs were carried out on the SSPCloud \citep{comte_sspcloud_2022}, a platform freely provided to French academics. This work made use of various open-source packages such as 
\texttt{numpy} \citep{harris_array_2020}
\texttt{matplotlib} \citep{hunter_matplotlib_2007},
\texttt{astropy} \citep{astropy_collaboration_astropy_2013, astropy_collaboration_astropy_2018,
astropy_collaboration_astropy_2022},
\texttt{ChainConsumer} \citep{hinton_chainconsumer_2016},
\texttt{cmasher} \citep{velden_cmasher_2020},
\texttt{haiku} \citep{hennigan_haiku_2020}.

\end{acknowledgements}

%
%

\bibliographystyle{aa} 
\bibliography{references.bib}

\appendix

\section{Basic cookbook}
\label{app:usecase}

The user can install \texttt{jaxspec} with a simple call to \texttt{pip}.

\begin{lstlisting}
pip install jaxspec
\end{lstlisting}

We recommand the user to start from a fresh \texttt{Python 3.10} or \texttt{3.11} environment. To build a spectral model, additive and multiplicative components can be tied together as follows : 

\begin{lstlisting}[language=Python]
from jaxspec.model.additive import Powerlaw
from jaxspec.model.additive import Blackbodyrad
from jaxspec.model.multiplicative import Tbabs 

model = Tbabs() * (Powerlaw() + Blackbodyrad())

\end{lstlisting}

If the data used are compliant with the OGIP standard, they can be loaded as follows : 

\begin{lstlisting}[language=Python]
from jaxspec.data import ObsConfiguration

obs = ObsConfiguration.from_pha_file(
    'path/to/data.pha',
    low_energy=0.5, high_energy=10
    # RMF and ARF can be explicitly stated
    #rmf_path='path/to/data.rmf', 
    #arf_path='path/to/data.arf',
)
    
\end{lstlisting}

Then, the data can be fitted by providing prior distributions.

\begin{lstlisting}[language=Python]
from jaxspec.fit import MCMCFitter
from numpyro.distributions import Uniform, LogUniform

prior = {
    "blackbodyrad_1_kT": Uniform(0, 3), 
    "blackbodyrad_1_norm": LogUniform(1e-3, 1),
    "powerlaw_1_alpha": Uniform(0, 3), 
    "powerlaw_1_norm": LogUniform(1e-5, 1e-1),
    "tbabs_1_N_H": Uniform(0, 1)
}

result = MCMCFitter(model, prior, obs).fit()
\end{lstlisting}

From this result, one can plot the posterior distributions, and perform posterior predictive checks. 

\begin{lstlisting}[language=Python]
result.plot_corner()
result.plot_ppc()
\end{lstlisting}

We refer the interested user to the documentation of \texttt{jaxspec}, which will evolve, unlike what is presented in this article. It can be found at \url{http://jaxspec.rtfd.io/}.

\section{Perseus model}
\label{app:perseus-model}

The unfolded spectrum is modelled by 9 Gaussian emission lines with a common broadening $\Sigma = \sigma_{v} / c$ and energy shift $\delta = v_{\text{bulk}}/c$. For each of these lines, we fit a free normalisation and an additional intrinsic energy perturbation, which should be much smaller than the total shift. 

\begin{equation}
    E_{L,i} = \frac{E_{\text{tab}, i}}{(1 + z)(1 + \delta)} + \Delta E_{\text{noise}, i}
\end{equation}

\noindent where $E_{\text{tab}, i}$ is the expected energy of line $i$ which can be found in the appendix of \cite{hitomi_collaboration_quiescent_2016}. This spectrum is then folded into \textit{Hitomi} responses for the three observations, for which we add a gain and shift parameter for two of them. An additional count-rate baseline is added for each instrument to model the continuum. The prior used for each parameter and the posterior values are displayed in Table \ref{tab:perseus-model}. We also use the automatic reparametrisation provided by \texttt{numpyro} to rescale and decenter the Gaussian prior distributions, and represent the log-uniform prior in a flat space, which regularise the likelihood and facilitates the gradient descent.

\renewcommand{\arraystretch}{1.5}\begin{table}[h]
\centering
\begin{tabular}{c|c|c}
Parameter & Prior distribution & Posterior 16-50-84 \% \\
\hline \hline
$\Sigma$ & $10^{\mathcal{U}(-5, -3)}$ & $6.54^{+0.12}_{- 0.12} x 10^{-4}$\\
$\delta$ & $\mathcal{N}(0, 10^{-3})$ & $1.70^{+0.49}_{- 0.48} x 10^{-4}$\\
$z$ & $\mathcal{N}(0.017284, 5\times 10^{-5})$ & $1.73^{+0.00}_{- 0.00} x 10^{-2}$\\
$\text{Norm}_0$ & $10^{\mathcal{U}(-5, -4)}$ & $1.57^{+0.22}_{- 0.16} x 10^{-5}$\\
$\text{Norm}_1$ & $10^{\mathcal{U}(-5, -4)}$ & $1.04^{+0.06}_{- 0.02} x 10^{-5}$\\
$\text{Norm}_2$ & $10^{\mathcal{U}(-5, -4)}$ & $9.51^{+0.20}_{- 0.32} x 10^{-5}$\\
$\text{Norm}_3$ & $10^{\mathcal{U}(-5, -4)}$ & $2.38^{+0.36}_{- 0.30} x 10^{-5}$\\
$\text{Norm}_4$ & $10^{\mathcal{U}(-5, -4)}$ & $3.67^{+0.35}_{- 0.32} x 10^{-5}$\\
$\text{Norm}_5$ & $10^{\mathcal{U}(-5, -4)}$ & $2.31^{+0.49}_{- 0.38} x 10^{-5}$\\
$\text{Norm}_6$ & $10^{\mathcal{U}(-5, -4)}$ & $7.76^{+0.39}_{- 0.44} x 10^{-5}$\\
$\text{Norm}_7$ & $10^{\mathcal{U}(-5, -4)}$ & $6.85^{+0.27}_{- 0.29} x 10^{-5}$\\
$\text{Norm}_8$ & $10^{\mathcal{U}(-5, -4)}$ & $2.49^{+0.05}_{- 0.04} x 10^{-4}$\\
$\Delta E_{\text{noise}, 0}$ & $\mathcal{N}(0, 10^{-4})$ & $-2.87^{+9.93}_{- 9.79} x 10^{-5}$\\
$\Delta E_{\text{noise}, 1}$ & $\mathcal{N}(0, 10^{-4})$ & $-0.40^{+9.79}_{- 9.78} x 10^{-5}$\\
$\Delta E_{\text{noise}, 2}$ & $\mathcal{N}(0, 10^{-4})$ & $1.98^{+0.95}_{- 0.94} x 10^{-4}$\\
$\Delta E_{\text{noise}, 3}$ & $\mathcal{N}(0, 10^{-4})$ & $-0.33^{+0.99}_{- 1.00} x 10^{-4}$\\
$\Delta E_{\text{noise}, 4}$ & $\mathcal{N}(0, 10^{-4})$ & $0.23^{+1.01}_{- 0.99} x 10^{-4}$\\
$\Delta E_{\text{noise}, 5}$ & $\mathcal{N}(0, 10^{-4})$ & $-0.13^{+1.01}_{- 1.00} x 10^{-4}$\\
$\Delta E_{\text{noise}, 6}$ & $\mathcal{N}(0, 10^{-4})$ & $-0.10^{+1.00}_{- 1.01} x 10^{-4}$\\
$\Delta E_{\text{noise}, 7}$ & $\mathcal{N}(0, 10^{-4})$ & $0.60^{+9.33}_{- 9.30} x 10^{-5}$\\
$\Delta E_{\text{noise}, 8}$ & $\mathcal{N}(0, 10^{-4})$ & $-1.53^{+0.96}_{- 0.95} x 10^{-4}$\\
$\text{Baseline (40)}$ & $\mathcal{U}(0, 10^{-3})$ & $6.37^{+0.39}_{- 0.40} x 10^{-4}$\\
$\text{Baseline (50)}$ & $\mathcal{U}(0, 10^{-3})$ & $6.76^{+1.07}_{- 1.26} x 10^{-4}$\\
$\text{Baseline (60)}$ & $\mathcal{U}(0, 10^{-3})$ & $8.29^{+0.46}_{- 0.60} x 10^{-4}$\\
$\text{Gain (50)}$ & $\mathcal{N}(1, 10^{-2})$ & $9.99^{+0.10}_{- 0.10} x 10^{-1}$\\
$\text{Gain (60)}$ & $\mathcal{N}(1, 10^{-2})$ & $1.01^{+0.01}_{- 0.01} x 10^{0}$\\
$\text{Shift (50)}$ & $\mathcal{N}(0, 10^{-3})$ & $-1.92^{+0.61}_{- 0.61} x 10^{-4}$\\
$\text{Shift (60)}$ & $\mathcal{N}(0, 10^{-3})$ & $-9.22^{+2.64}_{- 2.62} x 10^{-5}$\\
\hline
\end{tabular}
\caption{Prior distribution and posterior quantiles for the parameters of the Perseus core region model.}
\label{tab:perseus-model}
\end{table}

\end{document}